\documentclass[twocolumn,trackchanges]{aastex7}

\shorttitle{Fe K$\alpha$ line in Cygnus A}
\shortauthors{Majumder et al.}

\graphicspath{{./}{Figures/}}

\usepackage[T1]{fontenc}

\begin{document}

\title{XRISM detection of the 6.4 keV Fe K$\alpha$ line in the radio galaxy Cygnus A} 

\author[orcid=0000-0002-3525-7186]{Anwesh Majumder}
\affiliation{Waterloo Centre for Astrophysics, Department of Physics and Astronomy, 200 University Avenue West, Waterloo,  Ontario N2L 3G1, Canada}
\affiliation{Space Research Organisation Netherlands, Niels Bohrweg 4, Leiden, South Holland 2333 CA, The Netherlands}
\email[show]{anwesh.majumder@uwaterloo.ca}  

\author{T. Heckman}
\affiliation{Center for Astrophysical Sciences, William H. Miller III Department of Physics and Astronomy, Johns Hopkins University, Baltimore, Maryland 21218, USA}
\affiliation{School of Earth and Space Exploration, Arizona State University, Tempe, Arizona 85287, USA}
\email{thechma1@jhu.edu}

\author[orcid=0000-0001-9911-7038]{L. Gu}
\affiliation{Space Research Organisation Netherlands, Niels Bohrweg 4, Leiden, South Holland 2333 CA, The Netherlands}
\affiliation{Leiden Observatory, Leiden University, PO Box 9513, Leiden, South Holland 2300 RA, The Netherlands}
\affiliation{RIKEN High Energy Astrophysics Laboratory, 2-1 Hirosawa, Wako, Saitama 351-0198, Japan}
\affiliation{Department of Physics, Tokyo University of Science, 1-3 Kagurazaka, Shinjuku-ku, Tokyo 162-8601, Japan}
\email{l.gu@sron.nl}

\author[orcid=0000-0002-9714-3862]{A. Simionescu}
\affiliation{Space Research Organisation Netherlands, Niels Bohrweg 4, Leiden, South Holland 2333 CA, The Netherlands}
\affiliation{Leiden Observatory, Leiden University, Niels Bohrweg 2, Leiden, South Holland 2333 CA, The Netherlands}
\affiliation{Kavli Institute for the Physics and Mathematics of the Universe, The University of Tokyo, Kashiwa, Chiba 277-8583, Japan}
\email{a.simionescu@sron.nl}

\author{B.R. McNamara}
\affiliation{Waterloo Centre for Astrophysics, Department of Physics and Astronomy, 200 University Avenue West, Waterloo,  Ontario N2L 3G1, Canada}
\email{mcnamra@uwaterloo.ca}

\author[orcid=0000-0001-5655-1440]{A. Ptak}
\affiliation{NASA / Goddard Space Flight Center, 8800 Greenbelt Rd, Greenbelt, MD 20771, USA}
\affiliation{Center for Astrophysical Sciences, William H. Miller III Department of Physics and Astronomy, Johns Hopkins University, Baltimore, Maryland 21218, USA}
\email{ptak@pha.jhu.edu}

\author[orcid=0000-0002-2397-206X]{E. Hodges-Kluck}
\affiliation{NASA / Goddard Space Flight Center, 8800 Greenbelt Rd, Greenbelt, MD 20771, USA}
\email{edmund.hodges-kluck@nasa.gov}

\author[orcid=0000-0001-6366-3459]{M. Yukita}
\affiliation{NASA / Goddard Space Flight Center, 8800 Greenbelt Rd, Greenbelt, MD 20771, USA}
\affiliation{Center for Astrophysical Sciences, William H. Miller III Department of Physics and Astronomy, Johns Hopkins University, Baltimore, Maryland 21218, USA}
\email{myukita1@jhu.edu}

\author[orcid=0000-0002-6470-2285]{M.W. Wise}
\affiliation{Space Research Organisation Netherlands, Niels Bohrweg 4, Leiden, South Holland 2333 CA, The Netherlands}
\affiliation{Anton Pannekoek Institute, University of Amsterdam, Science Park 904, Amsterdam, North Holland 1098 XH, The Netherlands}
\email{m.w.wise@sron.nl}

\author[orcid=0000-0002-4430-8846]{N. Roy}
\affiliation{Center for Astrophysical Sciences, William H. Miller III Department of Physics and Astronomy, Johns Hopkins University, Baltimore, Maryland 21218, USA}
\affiliation{School of Earth and Space Exploration, Arizona State University, Tempe, Arizona 85287, USA}
\email{namratar@asu.edu}

\begin{abstract}
We detail the spectral analysis of a 170 ks XRISM \textit{Resolve} observation of the core of Cygnus A. The high spectral resolution of \textit{Resolve} have enabled us to probe the inner accretion region of Cygnus A by analyzing the 6.4 keV Fe K$\alpha$ line complex. We find that it consists of two Keplerian broadened components. (1) A broad component with a velocity dispersion of $3400^{+800}_{-600}$ km s$^{-1}$ and a narrow component of $440^{+60}_{-50}$ km s$^{-1}$. For an inclination of $50^{\circ}-85^{\circ}$, constrained by VLBI, we find that the broad component arises from a distance of $\sim 0.1-0.17$ pc ($800-1400$ gravitational radii) and the narrow component from  $\sim 6-10$ pc ($50,000-80,000$ gravitational radii) from the central black hole depending on the inclination angle. Our result suggests that the origin of the broad component is consistent with the broad line region and the narrow component from the torus of Cygnus A. We also find a potential emission line possibly from intermediate ionized Fe XVII with a very low dispersion ($<80$ km s$^{-1}$) that originates from either the outer edge of the torus or the narrow line region. Finally, we find that the Fe K edge is redshifted compared to the Fe K$\alpha$ line components, suggesting a line of sight bulk velocity of $470 \pm 100$ km s$^{-1}$. Such a shift may be due to an inflowing wind or relative motion between the two components originating from the near and far side of an inflowing torus, respectively.
\end{abstract}


\section{Introduction} 

Fe K$\alpha$ is a prominent atomic line feature in the X-ray spectrum commonly observed in non-blazar type active galactic nuclei (AGN; \citealt{pounds_x-ray_1990,1994MNRAS.268..405N}). The line arises due to the reflection of continuum hard X-rays from the corona off low- and intermediate-ionized Fe atoms in the circumnuclear environment surrounding the AGN. If the continuum radiation interacts with distinct parts of the accretion flow, then there can be multiple distinct components of this line. For instance, the Doppler effect can broaden lines originating from material closer to the black hole more than from material originating further away because of greater orbital speeds. The measured shape of these components can therefore trace the composition, geometry, and kinematics of the different parts of the accretion region. 

Before the advent of non-dispersive high-resolution X-ray spectroscopy, the Fe K$\alpha$ line was detected in X-ray bright Seyfert galaxies with dispersive gratings like High Energy Transmission Grating \citep{shu_cores_2010,shu_chandra_2011,gandhi_dust_2015,minezaki_new_2015}.  These past studies suggested that the line from Seyfert galaxies exhibits widths $> 2000$ km s$^{-1}$ and originates between the broad-line region (BLR) and dust reverberation radius. Recently, XRISM observed the Fe K$\alpha$ line in detail with its high spectral resolution capability ($\Delta E \sim 5$ eV at 6 keV; \citealt{2022SPIE12181E..1SI}) for the Seyfert galaxy NGC 4151 and revealed that the line actually has three distinct components that are associated with (a) the inner radii of torus (b) the inner radii of BLR and (c) a broad spectral component originating a few 100 gravitational radii from the central black hole \citep{xrism_collaboration_xrism_2024}. XRISM's recent success in resolving this line in such detail shows the power of high-resolution spectroscopy in constraining its physical origin. However, most past studies have focused on Seyfert galaxies and few on radio galaxies due to the lower corresponding equivalent widths and fluxes \citep{fukazawa_fe-k_2011}. NGC 1275 \citep{2018PASJ...70...13H} and Centaurus A \citep{bogensberger_xrism_2025} are the only radio-loud AGNs published so far at high spectral resolution in the X-rays. In NGC 1275, the origin of the Fe K$\alpha$ line can be as far out as the torus or molecular cloud ($\sim 1.6$ kpc), while in Centaurus A, the line shows different components arising between $10^{-3}-20$ pc. More high-resolution spectral work on radio-loud AGNs is required to infer whether the line's physical origin is similar or different from Seyfert galaxies.

In this work, we examine the powerful radio galaxy Cygnus A. Cygnus A is a prominent source that has helped to constrain models of active galactic nuclei (AGN) over time. At a redshift of $\sim 0.056$ \citep{1977ApJ...217...45S,owen_cluster_1997}, it is more than ten times more luminous than any other radio galaxy in the nearby universe \citep{1996A&ARv...7....1C}.  Cygnus A is an archetypal Fanaroff-Riley class II source (FR-II; \citealt{fanaroff_morphology_1974}) where the dominant radiation is from the lobes. The jet is driving a cocoon shock in the ambient medium around Cygnus A (e.g., \citealt{snios_cocoon_2018}) with a mechanical power of $\sim 10^{46}$ erg s$^{-1}$. At the center, the radio galaxy is powered by a type-II AGN \citep{antonucci_evidence_1994,1997ApJ...482L..37O}.  The nuclear region contains a supermassive black hole of $\sim 2.5 \times 10^9 M_{\odot}$ mass \citep{tadhunter_spectroscopy_2003}, estimated from HST and Keck spectroscopy of Pa-$\alpha$ and [O III] emission lines and under the assumption of circular motion. The region also contains a highly collimated jet originating from 200 times the Schwarschild radius \citep{boccardi_first_2016}, a radiative luminosity of $3 \times 10^{45}$ erg s$^{-1}$ \citep{privon_modeling_2012} and a neutral hydrogen absorber with a column density of $(2-4) \times 10^{23}$ cm$^{-2}$ \citep{ueno_x-ray_1994,young_chandra_2002,reynolds_nustar_2015}. The high column density obscures the inner accretion structure of Cygnus A in optical and ultraviolet. Thus, high spectral resolution X-ray analysis of reprocessed Fe Fluorescence can better shed insight into this part of the accretion region. Previously, this line was detected in Cygnus A using lower resolution \textit{Chandra} data \citep{young_chandra_2002} arising from $< 2$ kpc with an equivalent width of $182^{+40}_{-54}$ eV. In this work, we investigate the physical origin and kinematics of the components of this line in far greater detail with XRISM.


\begin{figure*}
    \centering
    \includegraphics[width=\textwidth, keepaspectratio]{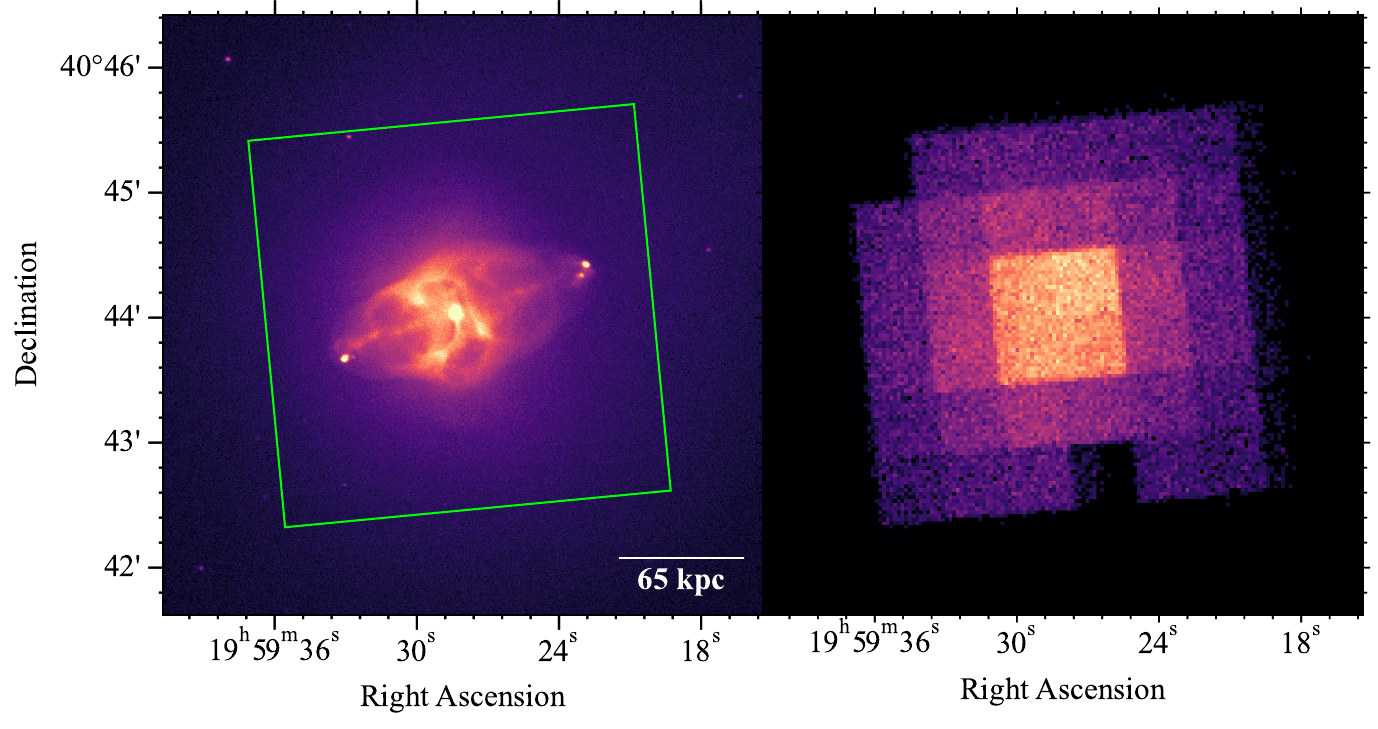}
    \caption{\small{
\emph{Left:} 2.2 Ms \textit{Chandra} image of the core of Cygnus A in the $0.7-7.0$ keV band. The image was processed according to the description in \cite{majumder_decoding_2024}. The \textit{Resolve} field of view has been overlaid in green. \emph{Right:} \textit{Resolve} view of the same region with pixels 12 and 27 removed. The image was created in the $1.7-12.0$ keV band. Only the Hp events were included.}}
\label{fig:fov}
\end{figure*}

We begin in \S\ref{sec:obs_data} with a description of the observation and the data reduction process. We then describe the spectral modeling in detail in \S\ref{sec:spec_model}. We present our results from the 6.4 keV line modeling in \S\ref{sec:results}. We conclude by discussing the implications of our work in \S\ref{sec:discussion}. The results reported in this work are a continuation of the work reported in \cite{majumder_spectrally_2026}. We used both cluster and AGN spectral models to accurately fit the \textit{Resolve} spectral data of Cygnus A. The results from the cluster model are reported in that work. In this work, we exclusively focus on the results from the AGN models. A $\Lambda$CDM cosmology with $H_0 = 70$ km s$^{-1}$ Mpc$^{-1}$, $\Omega_m = 0.3$ and $\Omega_{\Lambda} = 0.7$ have been assumed throughout this work. All errors reported in this work are of $1\sigma$ significance.

\section{Observation, Data Reduction and Spectral Extraction} \label{sec:obs_data}

The center of Cygnus A was observed for 170 ks between 6th October 2024 to 10th October 2024 (ObsID: 201120010) with an aimpoint of RA = $19^{\textrm{\scriptsize{h}}} 59^{\textrm{\scriptsize{m}}} 29^{\textrm{\scriptsize{s}}}.36$ and Dec = $40^{\circ} 44' 01''.6$ for a total net exposure of 170 ks. The data was reduced, and the full field of view spectrum was then extracted using the same procedure reported in \cite{majumder_spectrally_2026}. The full details are reported in that work. Here, we only briefly summarize the key information. The \textit{Resolve} field of view is shown in Figure \ref{fig:fov}.

The data was reprocessed using XRISM Build 8 software and CALDB v8, but with the latest RMF calibration file `\texttt{xa\_rsl\_rmfparam\_20190101v006.fits}'. The event file was filtered according to \texttt{RISE\_TIME} parameter by applying the criterion \texttt{((((RISE\_TIME+0.00075*DERIV\_MAX)>46) \
\&\&((RISE\_TIME+0.00075*DERIV\_MAX)<58))\&\&ITYPE<4)} \texttt{||(ITYPE==4))\&\&STATUS[4]==b0}. The spectrum was extracted from all pixels except pixel 27, which shows unexpected gain jumps. Only the High-Primary (highest energy resolution; Hp) events were retained. An X-sized redistribution matrix file (RMF) was then extracted and used in all our analyses as it models the secondary response components most accurately. A point source ancillary response file (ARF) was extracted to model the AGN emission, in addition to an extended ARF to model the extended cluster emission. The point source was centered at the aimpoint for the ARF generating task \texttt{xaarfgen}. The non-X-ray background (NXB) was extracted from a night-Earth database as recommended by the XRISM team\footnote{\href{https://heasarc.gsfc.nasa.gov/docs/xrism/analysis/nxb/resolve_nxb_db.html}{https://heasarc.gsfc.nasa.gov/docs/xrism/analysis/nxb/\\resolve\_nxb\_db.html}}. The NXB was then modeled using a power-law component along with Gaussian profiles for detector emission lines. This NXB model was used in all the following spectral fits.

\section{Spectrum preparation and modeling} \label{sec:spec_model}

\subsection{Spectrum Preparation for Fitting}

We used \texttt{SPEX v3.08.01} for all our spectral fitting \citep{1996uxsa.conf..411K,2018zndo...2419563K,2020zndo...4384188K}\footnote{\href{https://spex-xray.github.io/spex-help/index.html}{https://spex-xray.github.io/spex-help/index.html}}. As detailed in \cite{majumder_spectrally_2026}, all the spectral files were converted to the \texttt{SPEX} format, followed by optimally binning the response file \citep{kaastra_optimal_2016} in the $1.7-12.0$ keV band. With the help of the \texttt{SPEX} task \texttt{trafo}, we then arranged the data in two sectors (see \citealt{kaastra_optimal_2016} for a discussion on sectors) \footnote{\href{https://spex-xray.github.io/spex-help/theory/fitting/sectors.html}{https://spex-xray.github.io/spex-help/theory/fitting/sectors.html}}, where sector one contains the spectrum, RMF, and the extended source ARF, while sector two contains the same spectrum, RMF, but the point-source ARF. The cluster models were thus folded through the extended ARF and the AGN models were folded through the point source ARF. The extended ARF was created with the help of a \textit{Chandra} image in the $0.7-7.0$ keV band with the AGN removed. This arrangement of spectral files ensures that cluster and AGN models can be folded through their own respective ARFs.

\subsection{Spectral Modeling}

We only describe the AGN spectral models used to fit the $1.7-12.0$ keV spectrum in this work. In addition to these models, we also used spectral models to constrain the emission from the surrounding galaxy cluster. We refer the reader to \cite{majumder_spectrally_2026} for details on those models.

The shape of the 6.4 keV Fe K$_{\alpha}$ line was modeled with the laboratory measurement of the line \citep{1997PhRvA..56.4554H}. This data was also used for the analysis of the Fe fluorescence line in the NGC 4151 XRISM data \citep{xrism_collaboration_xrism_2024}. We loaded the data with the \texttt{file} model feature in \texttt{SPEX}. To model the 6.4 keV line accurately, we used two such \texttt{file} components, each being reflected and broadened from a different part of the accretion region. We allowed the normalizations ($N_{\textrm{\scriptsize{file,narrow}}}$ and $N_{\textrm{\scriptsize{file,broad}}}$) of the \texttt{file} components to vary and these \texttt{file} models were folded through the \texttt{SPEX} model \texttt{vgaus} for Keplerian broadening. All 6.4 keV components were then absorbed through an absorption column. The column density ($N_H$) was allowed to vary independently. The temperature of the \texttt{hot} model was fixed to $10^{-6}$ keV for neutral absorption. Finally, the 6.4 keV components were redshifted ($z_{\textrm{\scriptsize{6.4 keV}}}$) using the SPEX \texttt{reds} component. This redshift was allowed to vary to independently to determine the redshift of these components from the X-ray spectrum. The distance to the cluster in SPEX was set to $z=0.056$ \citep{1977ApJ...217...45S} and is consistent with line redshifts reported in Table \ref{tab:AGN_fits}.

We also find that the Fe K edge is at a higher redshift than that of the different redshift (see Section \ref{subsec:Fe_k-beta_edge}). We therefore modeled the non-thermal emission from the central AGN with a powerlaw (using the \texttt{pow} model in \texttt{SPEX}) and folded it through a different \texttt{hot} and redshift ($z_{\textrm{\scriptsize{edge}}}$) model. All the parameters of this \texttt{hot} model was coupled to the previously stated \texttt{hot} model, while the redshift was allowed to vary independently. The normalization and the powerlaw index of the powerlaw was allowed to vary independently as well.


For all our fits, we used the C-statistic for minimization \citep{1979ApJ...228..939C,kaastra_use_2017}. All abundances were measured with respect to the reference proto-solar abundance table of \cite{2009LanB...4B..712L}.

\section{Results} \label{sec:results}

\subsection{The 6.4 keV Fe emission line} \label{subsec:6.4 keV}

\begin{figure}
    \centering
    \includegraphics[width=\columnwidth, keepaspectratio]{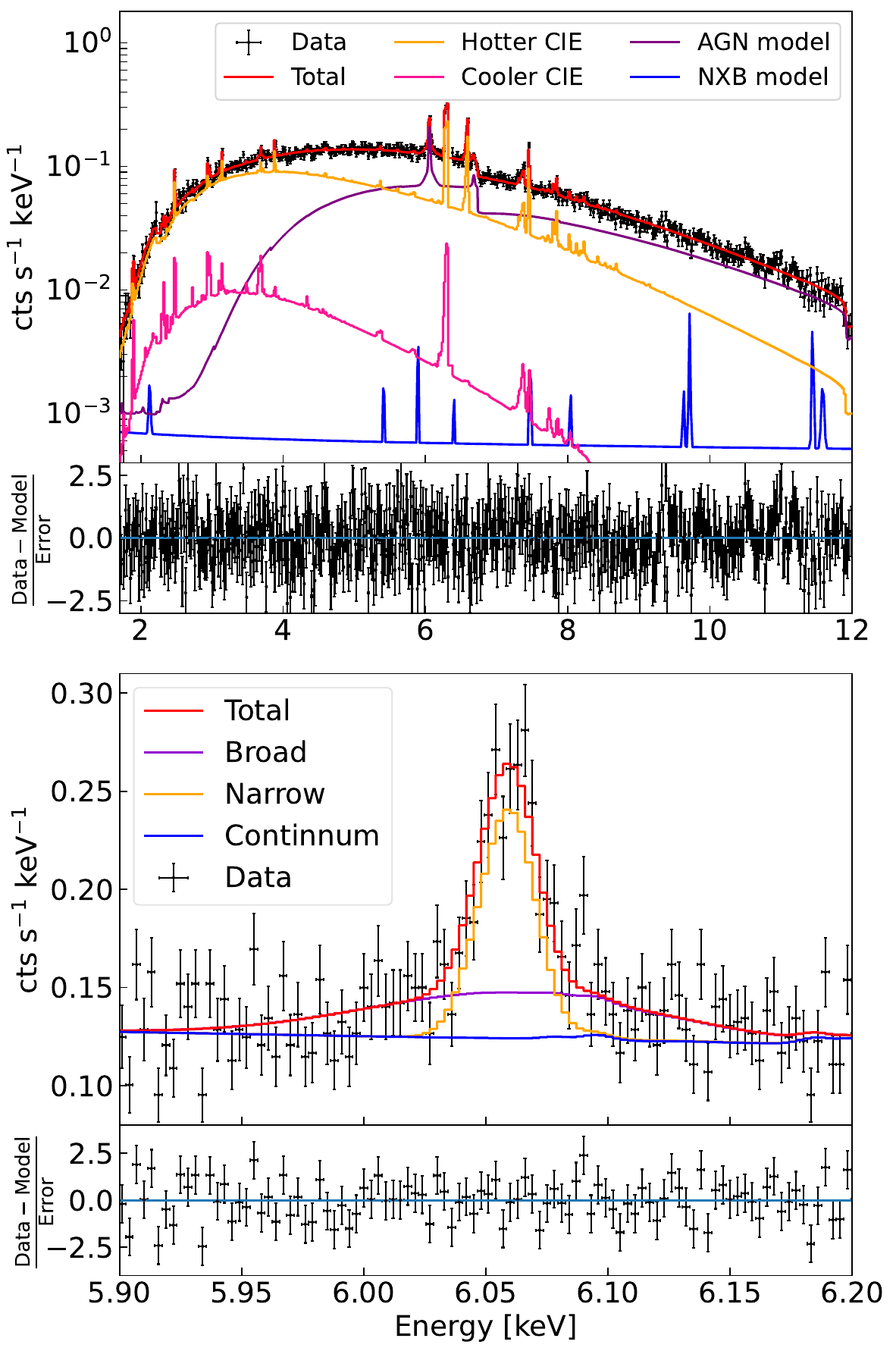}
    \caption{\small{Top:} The total AGN + intracluster medium (ICM) model fit to the $1.7-12.0$ keV band data, the individual model components (including hotter ICM, cooler ICM), along with the residuals as reported in \cite{majumder_spectrally_2026}. We also show the NXB model used for the fit. The spectrum has been binned by a factor of 5 in this energy range for visualization purposes. \emph{Bottom}: The 6.4 keV Fe line from the central AGN as measured by \textit{Resolve}. The narrow, broad, and total fitted model components are overplotted. The AGN powerlaw + cluster continuum is also shown for convenience.}
\label{fig:Fe_kalpha}
\end{figure}

The $1.7-12.0$ keV band was fitted with both AGN and cluster spectral models. The fitted spectrum results are shown in \ref{fig:Fe_kalpha}. The AGN spectral model parameters are reported in \ref{tab:AGN_fits}. The cluster spectral model parameters were previously reported in \cite{majumder_spectrally_2026}. 

Figure \ref{fig:Fe_kalpha} shows the energy band containing the 6.4 keV Fe fluorescence (redshifted to $\sim$6.06 keV) emission line and the spectral fit to it. The total fitted model, along with the reflected components from various parts of the accretion region, is overlaid. From Figure \ref{fig:Fe_kalpha}, it is clear that the 6.4 keV line consists of one narrow component and another much broader component. The velocities of these two components, other fit parameters, and the fit statistic are reported in Table \ref{tab:AGN_fits}. In \cite{majumder_spectrally_2026}, both one \texttt{cie} and two \texttt{cie} SPEX models were used to model the surrounding cluster emission. We find that the parameters of the AGN model are insensitive to the exact choice of the cluster model.

\begin{table}
\centering
\caption{Best-fit parameters of the AGN model for the Cygnus A \textit{Resolve} spectrum in the $1.7-12.0$ keV band.} 
\label{tab:AGN_fits}
    \begin{tabular}{cc}
    \hline
    \hline
    Parameter&Value\\
    \tableline
    \multicolumn{2}{c}{Neutral absorber} \\
         $N_H$ & $(4.0 \pm 0.2) \times 10^{23}$ cm$^{-2}$\\
    \tableline
    \multicolumn{2}{c}{Powerlaw component} \\
         $N_{\textrm{\scriptsize{pow}}}$ & $(2.7 \pm 0.4) \times 10^{9}$ \tablenotemark{a}\\
         $\Gamma$ & $2.07 \pm 0.06$\\
    \tableline
    \multicolumn{2}{c}{Narrow component} \\
        $z_{\textrm{\scriptsize{6.4 keV}}}$ &  $(5.596 \pm 0.018) \times 10^{-2}$\\
        $N_{\textrm{\scriptsize{file,narrow}}}$ & $3.7^{+0.6}_{-0.7} \times 10^4$\tablenotemark{a}\\
        $\sigma_{\textrm{\scriptsize{narrow}}}$ (km/s) & $440^{+100}_{-120}$\\
    \tableline
    \multicolumn{2}{c}{Broad component}\\
        $z_{\textrm{\scriptsize{6.4 keV}}}$ &  $(5.596 \pm 0.018) \times 10^{-2}$\\
        $N_{\textrm{\scriptsize{file,broad}}}$ & $(3.7 \pm 0.7) \times 10^4$\tablenotemark{a}\\
        $\sigma_{\textrm{\scriptsize{broad}}}$ (km/s) & $2900^{+1000}_{-800}$\\
    \tableline
    \multicolumn{2}{c}{Fe K edge}\\
        $z_{\textrm{\scriptsize{edge}}}$ & $(5.76 \pm 0.03) \times 10^{-2}$\\
    \tableline
    C-stat (Expected C-stat) & $3042$ ($3051 \pm 78$)\\
 \tableline
    \end{tabular}
    \tablecomments{The full model to fit the $1.7-12.0$ keV spectrum used was \texttt{reds(hot(pow))+reds(hot(vgaus(file)\\+vgaus(file)))} + cluster models. The cluster models are described in \cite{majumder_spectrally_2026}.}
    \tablenotetext{a}{The \texttt{file} and \texttt{pow} model normalizations have units of $10^{44}$ ph s$^{-1}$ keV$^{-1}$ at 1 keV.}
    
\end{table}

If we assume Keplerian motion for the clouds that produce the narrow and the broad component, we can derive a constrain a location following \cite{2026A&A...706A.255L}

\begin{equation} \label{eq:r_fe}
    R_{\textrm{\scriptsize{Fe}}} = \frac{GM_{\textrm{\scriptsize{BH}}}\textrm{sin}^2i }{5.5\sigma_{\textrm{\scriptsize{line}}}^2},
\end{equation}

where $M_{\textrm{\scriptsize{BH}}} = (2.5 \pm 0.7) \times 10^9 M_{\odot}$ is the central black hole mass \citep{tadhunter_spectroscopy_2003}, $\sigma_{\textrm{\scriptsize{line}}}$ is the measured Gaussian dispersion of the line components, $i$ is the inclination angle, and $G$ is the gravitational constant. This equation is very similar to that derived by \cite{2004ApJ...615..645O,2004ApJ...613..682P} from reverberation mapping but adjusted for geometric inclination. The inclination for Cygnus A was constrained between $50^{\circ}-85^{\circ}$ from Very Long Baseline Interferometry (VLBI) observations and modeling of the jet at pc-scale \citep{1996cyga.book...86S}.  In our work, we provide constraints on the radius using the VLBI limits on the inclination angle, although we do note that previous Spectral Energy Distribution (SED) analysis suggested an inclination closer to the higher end, i.e., $\sim 80^{\circ}$ \citep{privon_modeling_2012}. Using dispersions of the line components from Table \ref{tab:AGN_fits}, we derive $R_{\textrm{\scriptsize{Fe}}} \sim 0.14-0.23$ pc ($\sim 1160-1900$ gravitational radii) for the broad component and $\sim 6.7-11.3$ pc ($\sim 56,000-94,000$ gravitational radii) for the narrow component, assuming inclination angles between $50^{\circ}-85^{\circ}$.

\cite{tadhunter_spectroscopy_2003} suggested the size of the broad line region (BLR) in Cygnus A to be $\sim$$180$ lt-days = 0.2 pc from scaling relations of \cite{2000ApJ...533..631K}.  Furthermore, the inner walls of the torus in Cygnus A are expected to be at a distance of $0.3-0.8$ pc for a radiative luminosity of $3 \times 10^{45}$ erg s$^{-1}$ due to dust sublimation \citep{1987ApJ...320..537B,kishimoto_innermost_2007,burtscher_diversity_2013,netzer_revisiting_2015}. SED modeling by \cite{privon_modeling_2012} also suggests that the inner radius of the torus is likely $\sim$$0.6$ pc. The outer radius of the torus, on the other hand, is at a distance of $\sim 264$ pc \citep{carilli_imaging_2019}. Based on these numbers, our results suggest that the broad component of the 6.4 keV line arises from the BLR while the narrow component arises from the torus.

Although our fit was able to indicate the location of the components, the errorbars on the velocities are large. The derived errors on $R_{\textrm{\scriptsize{Fe}}}$ will thus be large as well. Most likely, this behavior is due to missing components that the error calculation from our fit is trying to compensate for. There are clear residuals at $\sim 6.08$ keV that needs to be modeled better. We thus consider an additional spectral model to better explain our data in Section \ref{subsec:Fe_Kalpha_3,4}.

We can also derive the relative velocity between the AGN (from $z_{\textrm{\scriptsize{AGN}}}$ of the 6.4 keV line) and the central galaxy ($z_{\textrm{\scriptsize{CG}}} = 0.05608 \pm 0.00007$; \citealt{owen_cluster_1997}) as

\begin{equation}   \label{eq:bulk_velocity}
    v_{\textrm{\scriptsize{bulk, AGN-CG}}} = \frac{c (z_{\textrm{\scriptsize{6.4 keV}}} - z_{\textrm{\scriptsize{CG}}})}{1 + z_{\textrm{\scriptsize{CG}}}},
\end{equation}

\noindent where $c$ is the speed of light. The redshifts suggest a relative velocity of $v_{\textrm{\scriptsize{bulk, AGN-CG}}} = -40 \pm 60$ km s$^{-1}$ (6.4 keV line components blueshifted). We therefore did not detect any systematic velocity offset between the AGN and the central galaxy. 

\subsection{A potential Fe XVII line} \label{subsec:Fe_Kalpha_3,4}

We model the line residual at $\sim 6.08$ keV with a delta line ($\texttt{delt}$ model in SPEX), broaden it with a \texttt{vagus} component, absorb with a neutral absorber and redshift it like the other 6.4 keV components. We allowed the normalization, energy of the \texttt{delt} model, and the Gaussian dispersion to vary independently. The flux of this line was measured to be $\sim 6-7\%$ of the narrow and broad components after fitting. Thus, this line can not be the Fe K$\alpha_{3,4}$ satellite line \citep{2006JPhB...39..651D}, which is at most a few percent of the main transition line. The most likely explanation of this line is due to emission from intermediate ionized Fe XVII ions \citep{2013PhRvL.111j3002R} with a rest energy of $\sim 6.435$ keV (redshifted to $\sim 6.09$ keV). Assuming photo-ionization, one can pre-dominantly produce Fe XVII emission by assuming a logarithmic ionization parameter (log $\xi$) of 1.8 in the \texttt{pion} model in SPEX \citep{2015Natur.526..542M,2016A&A...596A..65M}. We find that such an ionization parameter should also produce a weak Fe K$\beta$ emission with a rest energy of 7.2 keV (redshifted to $\sim 6.82$ keV). There are small residuals at $\sim 6.82$ keV in our data (see Figure \ref{fig:outflow}) but is difficult to discern due to the data quality. Due to data quality issues, we simply fit the residuals at $\sim 6.08$ keV with a broadened \texttt{delt} model rather than a dedicated photo-ionization mode like the \texttt{pion}.

\begin{figure}
    \centering
    \includegraphics[width=\columnwidth, keepaspectratio]{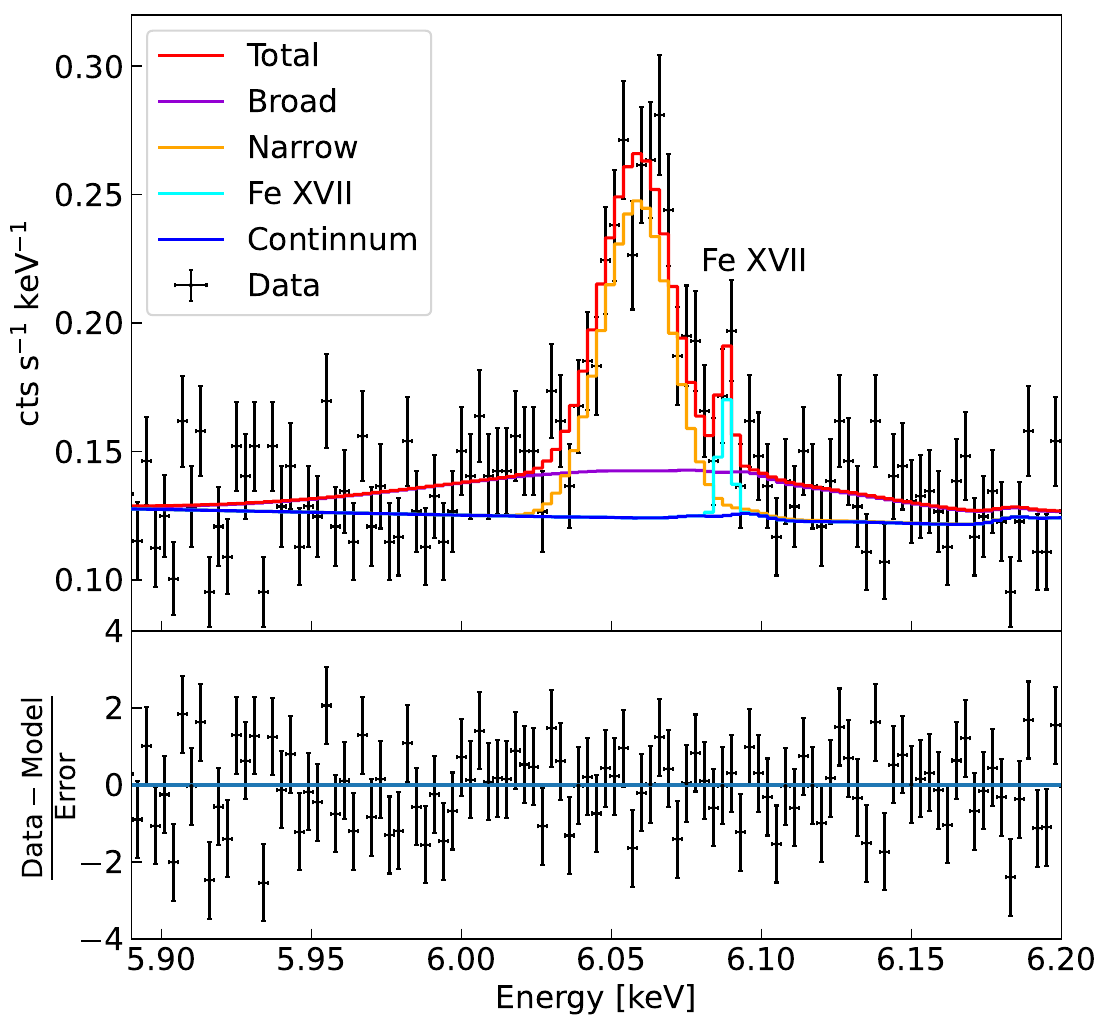}
    \caption{\small{\emph{Top:} Same as Figure \ref{fig:Fe_kalpha} but including a \texttt{delt} model broadened by a new \texttt{vgaus} model. The new models together provide a better fit to the 6.4 keV line complex. \emph{Bottom:} Residuals from the model fit.}}
\label{fig:Fe_kalpha_gaus}
\end{figure}

With these additional components, the spectrum was refit in the $1.7-12.0$ keV band. The fit is shown in Figure \ref{fig:Fe_kalpha_gaus}. As can be seen from Table \ref{tab:AGN_delt_fits}, the fractional velocity errors are smaller. Using Equation \ref{eq:r_fe}, we find $R_{\textrm{\scriptsize{narrow}}} = 6^{+3}_{-2}$ pc ($50,000^{+20,000}_{-17,000}$ gravitational radii) and $R_{\textrm{\scriptsize{broad}}} = 0.10^{+0.07}_{-0.04}$ pc ($800^{+500}_{-300}$ gravitational radii) for an inclination of $50^{\circ}$. Instead if the inclination is $85^{\circ}$, we find $R_{\textrm{\scriptsize{narrow}}} = 10^{+4}_{-3}$ pc ($80,000^{+30,000}_{-20,000}$ gravitational radii) and $R_{\textrm{\scriptsize{broad}}} = 0.17^{+0.11}_{-0.07}$ pc ($1,400^{+900}_{-600}$ gravitational radii). We used Monte Carlo sampling to calculate the median and the reported errors are 16th and 84th percentile of the distribution. Our results confirm that the spatial origin of the broad component is consistent with the BLR and the narrow component originates at the torus of Cygnus A. 

Finally, we can only place a lower limit on the distance of origin for the Fe XVII component (since $\sigma_{\textrm{\scriptsize{delt}}} < 80$ km s$^{-1}$). For $i = 50^{\circ}$, we obtain $R_{\textrm{\scriptsize{Fe XVII}}} > 190$ pc and $R_{\textrm{\scriptsize{Fe XVII}}} > 320$ pc for $i=85^{\circ}$. The outer radius of the torus is 264 pc and the height is 143 pc \citep{carilli_imaging_2019}. Assuming that this component is due to photo-ionization, it may originate from the outer edge of the torus. Alternatively, it can also arise from the Narrow Line Region (NLR) of Cygnus A. The NLR exists up to a radial distance of $1-2$ kpc \citep{2003MNRAS.342..995T,2025ApJ...983...98O}. The low dispersion is consistent with that from optical line measurements in the NLR with James Webb Space Telescope (see Figure 7 in \citealt{2025ApJ...983...98O}). We make further comments in Section \ref{sec:discussion}.

\begin{table}
\centering
\caption{Modified best-fit parameters of 6.4 keV Fe Fluorescence line after adding the \texttt{delt} model.} 
\label{tab:AGN_delt_fits}
    \begin{tabular}{cc}
    \hline
    \hline
    Parameter&Value\\
    \tableline
    \multicolumn{2}{c}{Narrow component} \\
        $N_{\textrm{\scriptsize{file,narrow}}}$ & $(3.9 \pm 0.2) \times 10^4$\tablenotemark{a}\\
        $\sigma_{\textrm{\scriptsize{narrow}}}$ (km/s) & $440^{+60}_{-50}$\\
    \tableline
    \multicolumn{2}{c}{Broad component} \\
        $N_{\textrm{\scriptsize{file,broad}}}$ & $(3.4 \pm 0.5) \times 10^4$\tablenotemark{a}\\
        $\sigma_{\textrm{\scriptsize{broad}}}$ (km/s) & $3400^{+800}_{-600}$\\
    \tableline
     \multicolumn{2}{c}{Potential Fe XVII K$\alpha$} \\
       $N_{\textrm{\scriptsize{\texttt{delt}}}}$ & $(2.2 \pm 0.8) \times 10^5$\tablenotemark{a}\\
        $E_{\textrm{\scriptsize{\texttt{delt}}}}$ (keV) & $6.430 \pm 0.002$\\
        $\sigma_{\textrm{\scriptsize{delt}}}$ (km/s) & $< 80$\\
    \tableline
    C-stat (Expected C-stat) & $3034$ ($3051 \pm 78$)\\
 \tableline
    \end{tabular}
    \tablecomments{The new model to fit the $1.7-12.0$ keV spectrum used was \texttt{reds(hot(pow))+reds(hot(vgaus(file)+\\vgaus(file)+vgaus(delt)))} + cluster models. Only the 6.4 keV line model parameters are noted in this table. Other unchanged AGN model parameters, are noted in Table \ref{tab:AGN_fits}.}
    \tablenotetext{a}{The \texttt{file} and \texttt{delt} model normalizations have units of $10^{44}$ ph s$^{-1}$ keV$^{-1}$ at 1 keV.}
    
\end{table}

The \texttt{cstat} of our new fit improves to 3034 ($\Delta C=8$). We can also calculate the Akaike information criterion (AIC) that is defined as

\begin{equation} \label{eq:aic}
 \textrm{AIC} = 2k - 2\textrm{ln}(\mathcal{L}),
\end{equation}

\noindent where $k$ is the number of parameters and $\mathcal{L}$ is the likelihood. C-stat is defined as $-2 \times$ Poissonian log-likelihood (see \citealt{1979ApJ...228..939C} and \citealt{kaastra_use_2017}). Equation \ref{eq:aic} then becomes:

\begin{equation}
    \textrm{AIC} = 2k + \textrm{C-stat}.
\end{equation}

For our previous fit (in Section \ref{subsec:6.4 keV}), $k=23$ and for our current fit, $k = 26$. Using the C-stat values from Tables \ref{tab:AGN_fits} and \ref{tab:AGN_delt_fits}, we obtain $\Delta$AIC = 2. The less complex model (in Section \ref{subsec:6.4 keV}) is 36.8\% as favored as the more complex model (Relative likelihood = $e^{-\Delta \textrm{\scriptsize{AIC}}/2}$). The more complex model is thus slightly favored (at 63.2\%). The more complex model also reduces the errorbars on dispersions of narrow and broad component. In this work, we report both models, noting that the more complex model is slightly favorable. However, we do emphasize that the present confidence level suggests that the feature may also be pure noise and hence we categorize it as only a potential Fe XVII line. A future deeper observation can confirm or rule out the existence of this feature.

\begin{figure}
    \centering
    \includegraphics[width=\columnwidth, keepaspectratio]{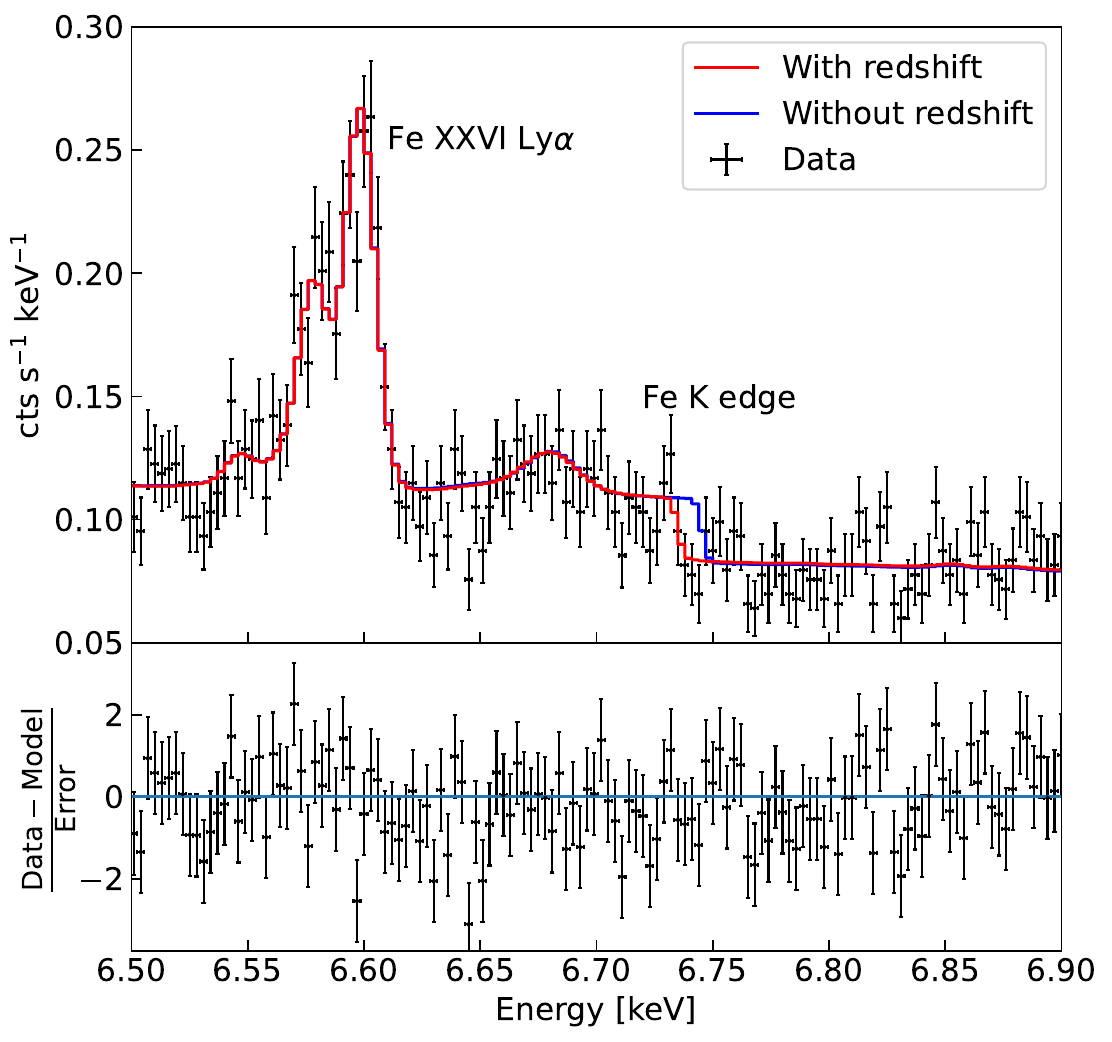}
    \caption{\small{\emph{Top:} \textit{Resolve} spectrum in the $6.5-6.9$ keV band along with the more redshifted Fe K edge fit. The model where the edge has same redshift as the 6.4 keV line complex is also shown to visualize the difference between the two. The Fe XXVI Ly$\alpha$ emission line originates from the cluster gas. \emph{Bottom:} Residuals from the more redshifted Fe K edge fit.}}
\label{fig:outflow}
\end{figure}

\subsection{Shift in Fe K edge} \label{subsec:Fe_k-beta_edge}

Next, we show the fitted spectrum in the $6.5-6.9$ keV band in Figure \ref{fig:outflow}, where the Fe K edge seems to be redshifted with respect to the 6.4 keV line components and the central galaxy. This effect is clear from Figure \ref{fig:outflow}. Without the extra redshift, the best fit C-stat was found to be 3047. When an extra redshift component is added, the C-stat improves to 3034 (Table \ref{tab:AGN_delt_fits}) and the AIC improves by 13. Such a large AIC overwhelmingly favors the addition of the extra redshift component. We also explored whether the edge can be modeled by an ionized absorber or by Doppler smoothening of line edges but found that the Fe K edge in our data is best modeled by assuming an extra redshift component.

Applying Equation \ref{eq:bulk_velocity}, we can calculate a bulk velocity of the absorber. We find a line of sight velocity of $v_{\textrm{\scriptsize{inflow,los}}} = 470 \pm 100$ km s$^{-1}$ with respect to the 6.4 keV emission regions (absorber redshifted). Since the redshift of the edge is different from the narrow and broad components of the 6.4 keV line complex, the substructure responsible for the Fe K edge shift is different from the 6.4 keV BLR or torus reflection regions. 

One simple interpretation that may explain such a redshifted bulk motion of the absorber is if there is an inflow. An inflow may occur if a wind fails to escape the gravitational potential well of the central AGN and falls back. If this explanation is accurate, the redshift of the edge with respect to the 6.4 keV line complex should change over time. Alternatively, the shift can also be caused if the torus as a whole is inflowing and the Fe K$\alpha$ is predominantly emitted by the far side of the torus, while the Fe K edge traces a line-of-sight absorber on the near side of the torus. The velocity of $470 \pm 100$ km s$^{-1}$ is then the relative velocity between the Fe K edge absorber and the 6.4 keV torus component. For edge-on configuration of torus, like that in Cygnus A, \cite{2021ApJ...913...17U} calculated that Fe K$\alpha$ may indeed be dominated by the far side of torus (see Figure 8). Therefore, such a geometric configuration may also sufficiently explain the relative observed shift between Fe K edge and Fe K$\alpha$ emission line components. 

We, however, do note that such a large shift is unlikely to be caused by atomic data calibration uncertainty in SPEX. The Fe K edge in \texttt{hot} model is at a rest energy of $\sim 7.123$ keV, consistent with the prominent K edge detected in Circinus X-1 by \textit{Resolve} (7.12 keV; \citealt{2025PASJ...77S..72T}), as modeled using \texttt{CLOUDY} \citep{2023RNAAS...7..246G,2025ApJ...991..203G} and the atomic database \texttt{chianti} \citep{2021ApJ...909...38D}. If the observed shift in K edge in Cygnus A was purely due to atomic data uncertainties, the rest frame energy would be $\sim 7.11$ keV. Such a large offset ($\sim$ 10 eV) would be inconsistent with the Circinus X-1 result. Thus we consider this scenario to be unlikely. 

\section{Discussion and Summary} \label{sec:discussion}

In our study, we demonstrated the power of high spectral resolution of XRISM \textit{Resolve} in isolating the physical origin of the 6.4 keV emission line from the nucleus of Cygnus A. While fitting the complex shape, we find that it is best fit by a broad Keplerian component that likely originates from the Broad Line Region (BLR) ($0.10^{+0.07}_{-0.04}$ pc = $800^{+600}_{-300}$ gravitational radii), and a narrow component that originates from the torus ($6^{+3}_{-2}$ pc = $50,000^{+20,000}_{-17,000}$ gravitational radii) for an assumed inclination of $50^{\circ}$. If instead, the inclination is changed to $85^{\circ}$, we obtain a distance of $0.17^{+0.11}_{-0.07}$ pc ($1,400^{+900}_{-600}$ gravitational radii) for the broad component and a distance of $10^{+4}_{-3}$ pc = $80,000^{+30,000}_{-20,000}$ gravitational radii). Our modeling on the \textit{Resolve} data, thus, has permitted exploration of inner accretion regions of Cygnus A. 

Previously, XRISM results from Seyfert 1 galaxies NGC 3783 \citep{2025A&A...699A.228M}, Mrk 279 \citep{2025ApJ...994L..10M}, Seyfert 1.5 galaxies NGC 4151 \citep{xrism_collaboration_xrism_2024}, NGC 3516 \citep{2025arXiv251207950J}, and Seyfert 2, FR-I \citep{fanaroff_morphology_1974} radio galaxy Centaurus A \citep{bogensberger_xrism_2025} also suggested that the 6.4 keV line complex is a combination of line features that originate from torus, BLR and sometimes from the accretion disk. From our work, we thus find that the Seyfert 2, FR-II radio galaxy Cygnus A's accretion structure is more similar to these galaxies, rather than radio-loud galaxies like NGC 1275, where the origin of the line is much further out ($\sim 1.6$ kpc; \citealt{2018PASJ...70...13H}). 

A potential Fe XVII K$\alpha$ emission line is also detected at an energy of $6.430 \pm 0.002$ keV with a flux of $\sim 6-7\%$ of the narrow and broad components. This line is narrow and only an upper limit could be placed ($\sigma_{\textrm{\scriptsize{delt}}} < 80$ km s$^{-1}$) and the line originates from $> 190$ pc for an inclination of $50^{\circ}$. If the inclination is increased to $85^{\circ}$, then the lower limit becomes 320 pc. As discussed in Section \ref{subsec:Fe_Kalpha_3,4}, the line can thus either originate from outer edges of the torus or the Narrow Line Region (NLR) of Cygnus A. With these lower limits on the distance of origin, it is possible to provide an upper limit on density using the relation $\xi = L/nr^2$, where $\xi$ is the ionization parameter, $L$ is the radiative luminosity, $n$ is the hydrogen density and $r$ is the distance from the ionizing source. Using $L = 3 \times 10^{45}$ erg s$^{-1}$ and log $\xi = 1.8$, we get $n < 140$ cm$^{-3}$ for $r > 190$ pc and $n < 50$ cm$^{-3}$ for $r > 320$ pc. It is interesting to note that these low densities are consistent with typical densities modeled in the NLR (\citealt{2025ApJ...983...98O}, also comparable to \citealt{2003MNRAS.342..995T}). A deeper future observation can help to get better statistics for this line, potentially better model it with photo-ionization models, obtain a more precise dispersion value and consequently constrain the spatial origin better. If this feature is real, there may also be contribution from Fe XVI and Fe XVIII, as well.

A Fe K edge redshift is also detected in the spectrum with respect to the 6.4 keV Fe K$\alpha$ components. As discussed in Section \ref{subsec:Fe_k-beta_edge}, the shift could simply be due to an inflowing wind or a relative motion between these two components, where the Fe K edge arises from the near side of the torus where it traces a line-of -sight absorber and the Fe K$\alpha$ components arise from the far side of an inflowing torus. For an edge-on torus like that in Cygnus A, Fe K$\alpha$ is expected to arise predominantly from the far side \citep{2021ApJ...913...17U}. Thus, such a dynamic configuration can also explain the observed shift in the Fe K edge. Such a shift, however, is unlikely to be entirely due to atomic data uncertainty in SPEX as the observed rest frame energy ($\sim 7.11$ keV) of the edge will be inconsistent with the prominent Fe K edge detection in Circinus X-1 (rest energy 7.12 keV; \citealt{2025PASJ...77S..72T}). 

We did not find compelling evidence of a Compton shoulder that should appear as a broad line shoulder at the lower energies. There are some small positive residuals between $6.00-6.03$ keV (see Figure \ref{fig:Fe_kalpha_gaus}), but the improvement in fit statistic when fitted with an additional Compton shoulder model (like \texttt{vcom} model in SPEX) is negligible. Any presence of the Compton shoulder, thus, could not be resolved from the broad line component that extends towards both higher and lower energies. Much deeper exposure may be required to investigate any presence of Compton shoulder.
 
\begin{acknowledgments}
We thank the anonymous referee for critical comments that led to substantial improvement of the manuscript. TH acknowledges the financial support from NASA 80NSSC25K7537 for this project. AM, AS, LG, and MW acknowledge support from the Netherlands Organisation for Scientific Research (NWO). BRM acknowledges support from the Canadian Space Agency and the National Science and Engineering Research Council of Canada. This research has made use of data obtained from the XRISM data archive maintained by NASA HEASARC and JAXA DARTS. The research also made use of the \textit{Chandra} data archive provided by the \textit{Chandra} X-ray Center (CXC). 
\end{acknowledgments}

\section*{Software and third party data repository citations}

The XRISM raw data used in this work is publicly available at NASA HEASARC and JAXA DARTS. The Chandra datasets used to create Figure \ref{fig:fov} are contained in the DOI ~\dataset[https://doi.org/10.25574/cdc.605].. All intermediate data products and scripts used to obtain the results are publicly available at \cite{maj25-zen}.

\facilities{XRISM (\textit{Resolve}), \textit{Chandra} (ACIS)}

\software{\texttt{NUMPY} \citep{2020NumPy-Array}, 
          \texttt{ASTROPY} \citep{astropy:2013,astropy:2018,astropy:2022},
          \texttt{MATPLOTLIB} \citep{hun07},
          \texttt{APLPY} \citep{rob12}}

\pagebreak


\end{document}